\documentclass[10pt,aps,prl,twocolumn,superscriptaddress,floatfix]{revtex4-2}
\usepackage{graphicx}%
\usepackage{multirow}%
\usepackage{amsmath,amssymb,amsfonts}%
\usepackage{amsthm}%
\usepackage{mathrsfs}%
\usepackage{xcolor}%
\usepackage{textcomp}%
\usepackage{booktabs}%
\usepackage{bm}%
\usepackage{amsmath}
\usepackage{siunitx}

\begin{document}

\title{Element-Specific Visualization of Layer-Parity and Twist-Dependent Magnetism in CrSBr}

\author{Aalok Tiwari}
\altaffiliation{These authors contributed equally to this work.}
\affiliation{Department of Physics, Carnegie Mellon University, Pittsburgh, PA~15213, USA}

\author{Shubhada Patil}
\altaffiliation{These authors contributed equally to this work.}
\affiliation{Helmholtz-Zentrum Berlin f\"{u}r Materialien und Energie,
  Albert-Einstein-Stra\ss{}e~15, 12489~Berlin, Germany}

\author{Ravi Kumar Bandapelli}
\altaffiliation{These authors contributed equally to this work.}
\affiliation{Department of Physics, Carnegie Mellon University, Pittsburgh, PA~15213, USA}

\author{Alevtina Smekhova}
\affiliation{Helmholtz-Zentrum Berlin f\"{u}r Materialien und Energie,
Albert-Einstein-Stra\ss{}e~15, 12489~Berlin, Germany}

\author{Abhishek Kumar}
\affiliation{Computational Materials Physics Laboratory, Department of Physics, Indian Institute of Technology Ropar, Rupnagar, Punjab 140001, India}

\author{Wenhao Liu}
\affiliation{Department of Physics, The University of Texas at Dallas,
  Richardson, TX~75080, USA}

\author{I-Hsuan Kao}
\affiliation{Department of Physics, Carnegie Mellon University, Pittsburgh, PA~15213, USA}

\author{Zhenhong Cui}
\affiliation{Department of Physics, Carnegie Mellon University, Pittsburgh, PA~15213, USA}

\author{Raghvendra Posti}
\affiliation{Department of Physics, Carnegie Mellon University, Pittsburgh, PA~15213, USA}

\author{Brandon Tran}
\affiliation{Department of Physics, Carnegie Mellon University, Pittsburgh, PA~15213, USA}

\author{Zixin Zhai}
\affiliation{Department of Physics, The University of Texas at Dallas,
  Richardson, TX~75080, USA}

\author{Priti Yadav}
\affiliation{Department of Physics, The University of Texas at Dallas,
  Richardson, TX~75080, USA}

\author{Sandy Adhitia Ekahana}
\affiliation{Department of Physics, Carnegie Mellon University, Pittsburgh, PA~15213, USA}

\author{Alexander X. Gray}
\affiliation{Department of Physics, Temple University,
  Philadelphia, PA~19122, USA}

\author{Bing Lv}
\affiliation{Department of Physics, The University of Texas at Dallas, Richardson, TX~75080, USA}
\affiliation{Department of Materials Science \& Engineering, The University of Texas at Dallas,
Richardson, TX 75080 USA}
  
\author{Vivekanand Shukla}
\affiliation{Computational Materials Physics Laboratory, Department of Physics, Indian Institute of Technology Ropar, Rupnagar, Punjab 140001, India}

\author{Florian Kronast}
\affiliation{Helmholtz-Zentrum Berlin f\"{u}r Materialien und Energie,
  Albert-Einstein-Stra\ss{}e~15, 12489~Berlin, Germany}
  
\author{Simranjeet Singh}
\email{simranjs@andrew.cmu.edu}
\affiliation{Department of Physics, Carnegie Mellon University, Pittsburgh, PA~15213, USA}

\author{Jyoti Katoch}
\email{jkatoch@andrew.cmu.edu}
\affiliation{Department of Physics, Carnegie Mellon University, Pittsburgh, PA~15213, USA}

\date{\today}

\begin{abstract}
Van der Waals (vdW) based antiferromagnets (AFMs) are an ideal platform for probing and understanding thickness- and twist-angle-dependent emergent spin phenomena. However, element-specific nanoscale characterization of the spin structure in atomically thin vdW-based AFMs systems and layer-parity effects remain elusive, making them crucial for both fundamental insight into low-dimensional magnetism and the rational design of spintronic devices based on these materials. Here, we utilize X-ray magnetic circular and linear dichroisms paired with photoemission electron microscopy to resolve the magnetic order in atomically thin CrSBr. Our comprehensive measurements reveal CrSBr magnetic structure at the nanoscale and its dependence on the layer number, surface encapsulation, temperature, and applied field. Moreover, in the orthogonally twisted bilayer configuration, obtained by twisting two CrSBr ferromagnetic monolayers by 90$^\circ$, the magnetic easy axis fundamentally differs from the individual monolayers, unlocking a new pathway for moir\'e magnetism.
\end{abstract}

\maketitle

\section*{Introduction\label{sec:intro}}

Discovery of intrinsic magnetism in atomically thin van der Waals (vdW) materials has opened new avenues for understanding and manipulating spin order in low-dimensional systems \cite{burch2018magnetism,mak2019probing,gibertini2019magnetic}. Specifically, vdW-based systems offer a unique material platform to develop a comprehensive understanding of static and dynamical spin phenomena in antiferromagnets (AFMs) because of their tunable properties  \cite{yang2021van, zhou2022dynamical, xu2025magnetostatic}. Furthermore, atomically precise assembly of vdW-based magnetic heterostructures offers a modular route to explore emergent magnetic phenomena with promise for spintronic applications ranging from sensing to data storage \cite{soumyanarayanan2016emergent,gong2017discovery,song2021direct}. Recent studies suggest that twist engineering of magnetic layers generates moir\'e magnetic exchange interactions and stabilizes exotic phenomena including coexisting ferromagnetic-antiferromagnetic (FM-AFM) states in bilayer CrI$_3$ \cite{yang2023moire,xu2022coexisting,jang2024direct}. Additionally, noncollinear magnetic states, topological skyrmion lattices, strongly bounded excitons and electrically tunable moir\'e magnetism have been predicted theoretically and realized experimentally \cite{tong2018skyrmions,hejazi2020noncollinear,wang2020stacking,akram2021skyrmions,cheng2023electrically,li2024observation,adak_excitons_2026,kim2026emergent}. 

Among layered AFMs, CrSBr has emerged as an exceptional platform combining air stability, an accessible N\'eel temperature ($T_{\mathrm{N}} = 132$~K), and a quasi-one-dimensional semiconducting electronic structure  \cite{telford2020layered,lee2021magnetic,liu2022three,wilson2021interlayer,klein2023bulk}. CrSBr crystallizes in the orthorhombic space group \textit{Pmmn} ($D^{13}_{2h}$) with point symmetry $mm2$ ($C_{2v}$) at Cr sites \cite{goser1990magnetic}. In bulk form, it adopts A-type AFM ordering, i.e., individual layers exhibit internal FM alignment with strong intralayer exchange, while adjacent layers couple antiferromagnetically through much weaker interlayer exchange \cite{brennan2024important}. Pronounced triaxial magnetic anisotropy with easy, intermediate, and hard axes along the crystallographic $b$, $a$, and $c$ directions arises from spin orbit coupling at both Cr and Br sites \cite{wang2023magnetic}. The unique combination of semiconducting character and weak interlayer coupling has enabled observations of magnon-exciton interactions, magneto-electric phenomena, and magnon-mediated spin transport \cite{bae2022exciton,wu2022quasi,datta2025magnon,liebich2025controlling}. Twist engineering of bilayers has been utilized to construct all-AFM tunnel junctions \cite{chen2024twist}, while multistep magnetization switching in orthogonally twisted bilayers has been reported \cite{boix2024multistep}. First principle calculation suggests a dramatic transition in magnetic anisotropy at specific angles \cite{zhang2025twist}. 

The comprehensive understanding of evolution of magnetic structure from the few-layer limit to twisted regime and the effects of external perturbations is crucial for both fundamental insight into low-dimensional magnetism and the rational design of vdW-based spintronic devices. Although the local magnetometry measurements \cite{thiel2019probing,rizzo2022visualizing,ghiasi2023nitrogen,zur2023magnetic,tschudin2024imaging} have provided both quantitative and qualitative understanding of a few-layer CrSBr magnetization, yet element-specific characterization of the Cr spin structure down to monolayer and twisted regime remains elusive. 

\begin{figure*}[!ht]
  \includegraphics[width=0.8\linewidth]{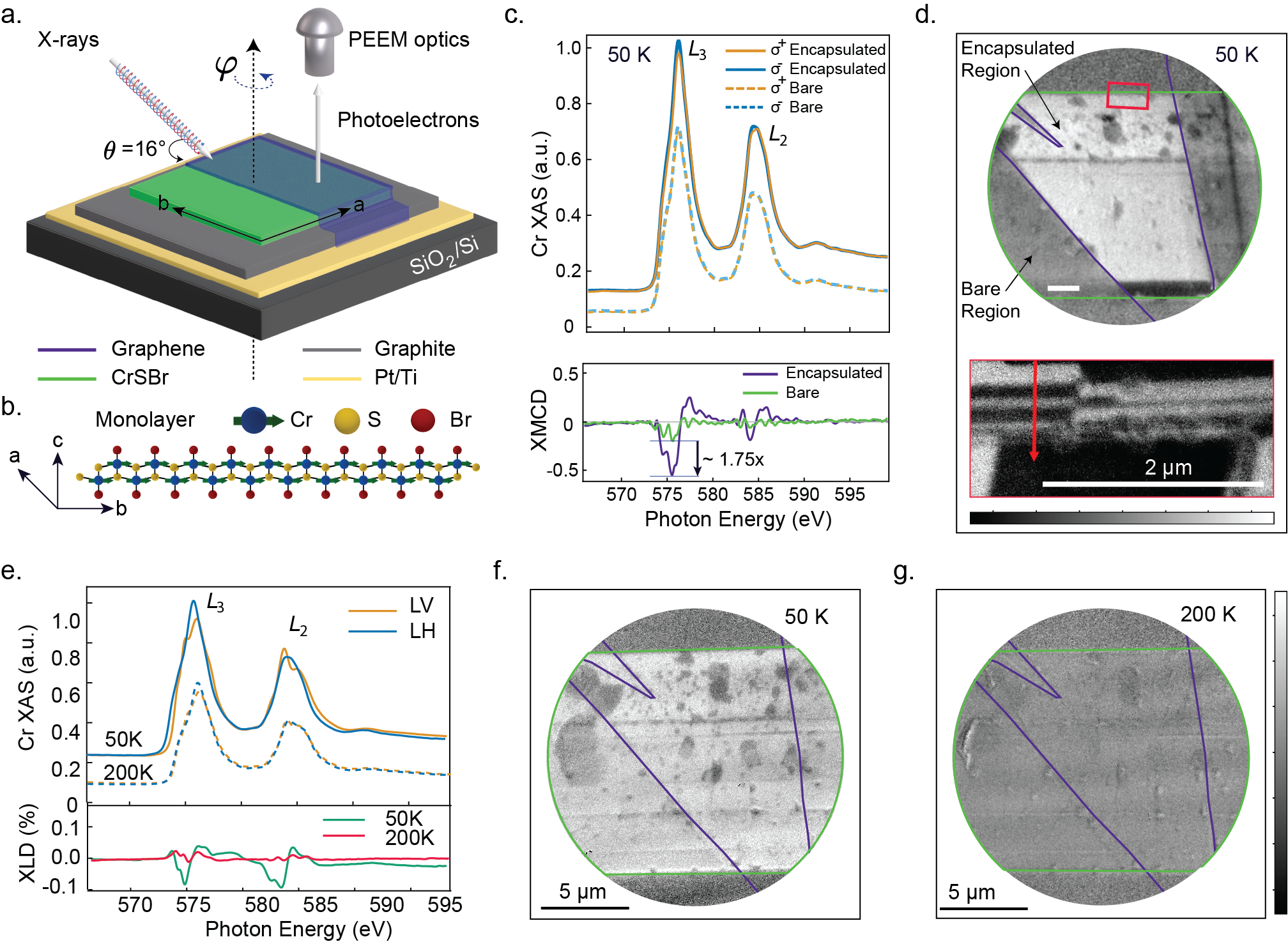}
  \caption{\label{Figure1}
  \textbf{Graphene encapsulation enhances XMCD signal in CrSBr.}
  \textbf{a}, Schematic of PEEM experimental geometry with X-rays at $16^{\circ}$ grazing incidence and azimuthal rotation ($\varphi$) capability. The heterostructure consists of CrSBr (green) encapsulated between graphene (purple) and graphite (gray) on a Pt/Ti/SiO$_{2}$/Si substrate.
  \textbf{b}, Atomic structure of monolayer CrSBr showing Cr (blue), S (red), and Br (yellow) atoms. Cr magnetic moments are aligned along the crystallographic $b$-axis.
  \textbf{c}, X-ray absorption spectra (XAS-PEEM) at the Cr $L_{2,3}$ edges acquired with right ($\sigma^{+}$, orange) and left ($\sigma^{-}$, blue) circularly polarized photons, comparing graphene-encapsulated (top) and bare (bottom) CrSBr regions at 50~K. Lower panel: XMCD asymmetry showing enhanced dichroic signal in graphene-encapsulated regions.
  \textbf{d}, XMCD-PEEM map at 50~K of the Cr $L_{3}$ edge in a multilayer CrSBr flake (upper panel). Inset (red box): magnified view of surface-termination-dependent net magnetization. Black regions: magnetization parallel to $b$-axis; white regions: magnetization antiparallel. Scale bar: 2~$\mu$m.
  \textbf{e}, XAS-PEEM spectra at the Cr $L_{2,3}$ edges acquired with linear vertical (LV, orange) and linear horizontal (LH, blue) polarized photons, measured above (200~K) and below (50~K) the bulk Néel temperature in the graphene-encapsulated region. Lower panel showing significantly reduced XLD contrast at 200~K.
 XLD-PEEM map at \textbf{f} 50~K and, \textbf{g}, 200~K. 
 showing reduced linear dichroism above the Néel temperature.   The observed XLD contrast primarily arises from crystal field effects (CFEs).
  Green outlines mark CrSBr regions; violet outlines indicate graphene in all PEEM maps. Grayscale colorbars indicate normalized dichroic contrast intensity. Scale bars: 5~$\mu$m.}
\end{figure*}

For this, X-ray linear and magnetic circular dichroism (XLD and XMCD) spectroscopy paired with photoemission electron microscopy (PEEM) can provide element specific sensitivity to magnetic moments and Néel vector mapping  with nanoscale spatial resolution \cite{bruche1933elektronenmikroskopische,stohr1998principles,stohr2006magnetism,vaz2025x}. Here, we employ XMCD/XLD combined with PEEM to directly visualize element-specific magnetic domain morphology from monolayer to tetralayer CrSBr with nanoscale spatial resolution. Complementing stray-field and local magnetometry techniques, XMCD-PEEM provides direct sensitivity to Cr magnetic moments and their alignments, enabling holistic characterization of domain structure and interlayer coupling. We demonstrate layer-parity-dependent spin structure (odd versus even layers) consistent with A-type AFM order from multilayer to monolayer limit. We further manipulate the observed spin structure using in-situ, in-plane external magnetic field.

\section*{Results\label{Resultss}}

\begin{figure*}[ht]
  \includegraphics[width=0.85\linewidth]{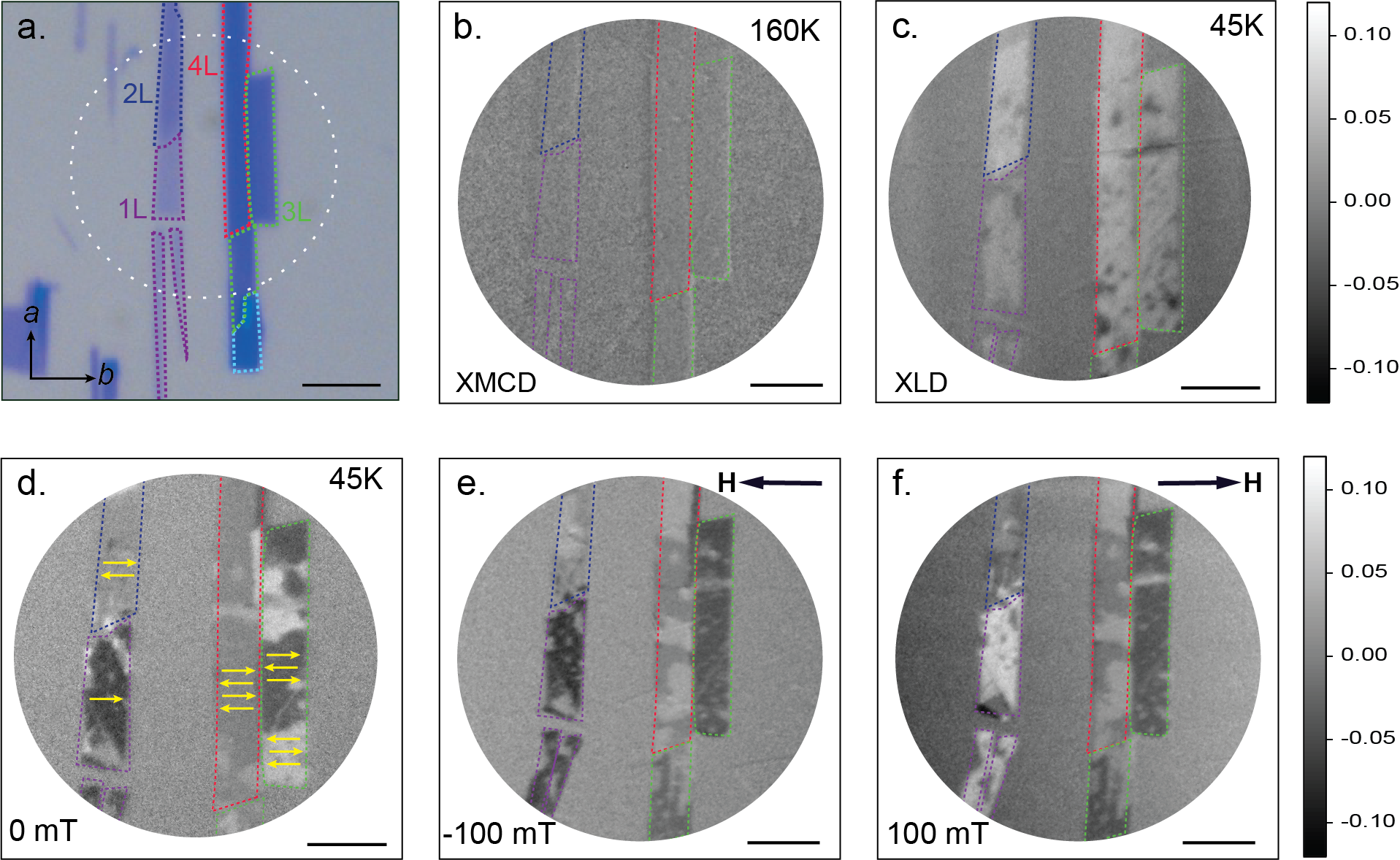}
  \caption{\label{Figure2}
  \textbf{Layer-parity-dependent magnetic structure in atomically thin (1L--4L) CrSBr.}
  \textbf{a}, Optical micrograph of exfoliated CrSBr showing regions of increasing thickness: 1L (violet), 2L (blue), 3L (green), and 4L (red). Scale bar: 5~$\mu$m.
  \textbf{b}, XMCD-PEEM map acquired at 160~K (above the bulk Néel temperature), showing absence of magnetic dichroic contrast. 
  \textbf{c}, XLD-PEEM map at 45~K revealing finite linear dichroic contrast across all layers.
  \textbf{d}, Ground-state XMCD-PEEM map at 45~K showing layer-parity-dependent magnetization: odd-numbered layers (1L, 3L) exhibit net magnetization (white/black contrast), while even-numbered layers (2L, 4L) remain magnetically compensated (gray).
  \textbf{e}, XMCD-PEEM map of the remanent magnetization state after applying a magnetic field pulse of $-$100~mT along the easy axis.
  \textbf{f}, Remanent state following a reversal field pulse of $+$100~mT, demonstrating field-induced magnetization switching in odd layers while even layers remain unaffected.
  Colored dashed outlines delineate layer boundaries. Scale bars: 2~$\mu$m for all PEEM images. Grayscale colorbars indicate dichroic contrast intensity.}
\end{figure*}

Exfoliated CrSBr flakes with varying thicknesses are placed on a thick graphite and are covered with monolayer graphene (Fig.~1a), prepared using dry mechanical transfer technique using a custom-built transfer tool in an argon-filled glove box  \cite{zomer2014fast,kao2022deterministic,ekahana2026visualization}. All X-ray measurements were performed with circularly or linearly polarized soft X-rays incident at a grazing angle of $16^{\circ}$ relative to the sample surface along the magnetic easy axis. Fig.~1b shows monolayer CrSBr with spins aligned along magnetic easy axis (crystallographic b-axis). Photoelectrons were collected by a PEEM that includes energy analyzer, enabling element specific spectro-microscopy with nanoscale spatial resolution. First, we characterize a representative thick CrSBr sample comprising regions with and without graphene encapsulation (H01). X-ray absorption spectroscopy (XAS) at the Cr $L_{2,3}$ edge reveals well-defined multiplet features and sizable XMCD below the magnetic ordering temperature. Graphene encapsulated regions display enhanced Cr signal and larger XMCD asymmetry compared to non encapsulated (bare) regions, confirming suppression of surface degradation and enhanced surface quality (Fig.~1c,d).

The net magnetization direction in the thicker samples is determined by the surface termination, i.e., whether the topmost layer is odd or even (Fig.~1d, lower panel).  Importantly, this behavior remains unchanged upon field cooling. The topmost layer spin orientation dominates the depth-averaged XMCD signal, yielding magnetization polarized either parallel or antiparallel to the easy axis depending on the layer number. The corresponding AFM topography and height profile are shown in Supplemental Information, Fig.~S1b,c and Table.~S1. This layer-parity-dependent alternating pattern is intrinsic to the A-type AFM order in bulk CrSBr.

The finite X-ray linear dichroism (XLD) persists from 50~K up to 200~K. The XLD signal comprises two contributions: X-ray natural linear dichroism (XNLD), arising from crystal field effects and $d$-orbital anisotropy, and XMLD, which reflects magnetic anisotropy. At 50~K, both XNLD and XMLD contributions are present across the multilayer regions (Fig.~1e--f). The XMLD dominates below the Néel temperature and is directly related to the preferred in-plane orientation of the easy axis ($\vec{b}$-direction) in the multilayer system. In contrast, at 200~K, which is above the magnetic ordering temperature, the persistent non-vanishing XNLD signal (Fig.~1f) is attributed to crystal field effects arising from Cr $d$-orbital anisotropy \cite{pei2024surface,poree2025resonant}.

\begin{figure*}[ht]
  \includegraphics[width=0.85\linewidth]{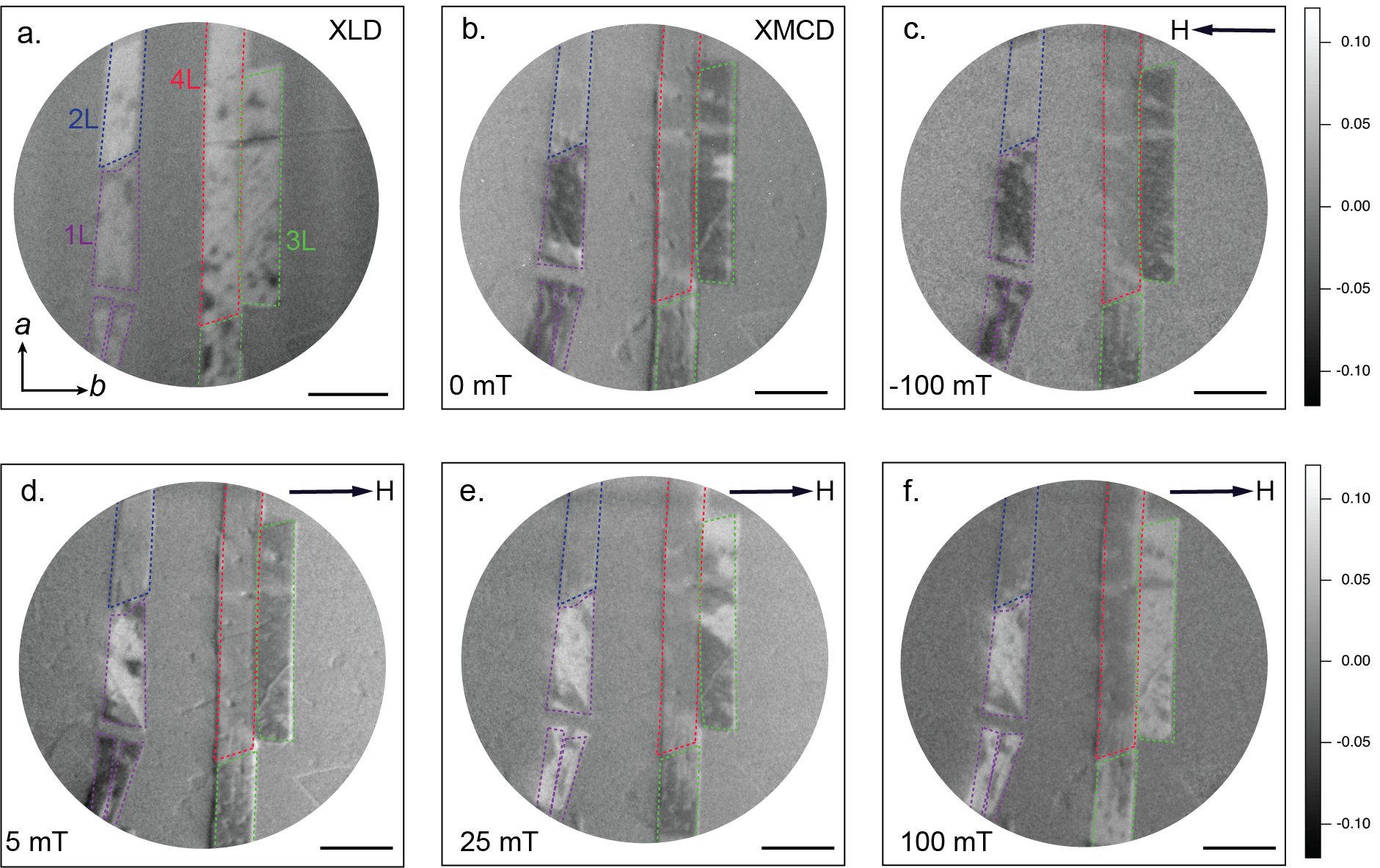}
  \caption{\label{Figure3}
  \textbf{Magnetic field-driven spin structure in atomically thin CrSBr at 110~K.} The sample was field-cooled from the paramagnetic state through the magnetic transition temperature under an applied field of $-$25~mT to initialize domain alignment. Colored dashed lines delineate layer boundaries (1L, violet; 2L, blue; 3L, green; 4L, red).
  \textbf{a}, XLD-PEEM map showing finite linear dichroic contrast across all layers.
  \textbf{b}, XMCD-PEEM map acquired after field-cooling in $-$25~mT, revealing preferential alignment of odd-numbered layers along the field direction.
  \textbf{c}, Application of $-$100~mT saturates the net magnetization in odd layers fully along the field direction; even layers remain compensated.
  \textbf{d}, Under a weak reversal field of $+$5~mT, the 1L magnetization nearly reverses, while the 3L remains largely unchanged, demonstrating layer-dependent magnetic response.
  \textbf{e}, At $+$25~mT, the 1L magnetization fully reverses; the 3L exhibits partial domain switching with two coexisting magnetization orientations.
  \textbf{f}, At $+$100~mT, all odd-layer regions achieve complete magnetization reversal, while even-layer regions maintain their compensated AFM state without detectable change. Scale bars: 2~$\mu$m. Grayscale colorbars indicate dichroic contrast intensity.}
\end{figure*}

\subsection*{Layer parity dependent magnetism in atomically thin limit}

Next, we examine atomically thin CrSBr layers (from 1L to 4L), within the same sample (H02), to establish layer-dependent magnetism persisting down to ultrathin limit (Fig.~2a). Temperature dependent measurements reveal that XMCD contrast is absent above the bulk Néel temperature ($T_N = 132$~K, Fig.~2b), while both significant XLD (Fig.~2c) and XMCD (Fig.~2d) contrast is observed at 45~K across all layer thicknesses. The observed dichroism is separated by sharp domain boundaries across all layer thicknesses. 

\begin{figure*}[ht]
  \includegraphics[width=0.85\linewidth]{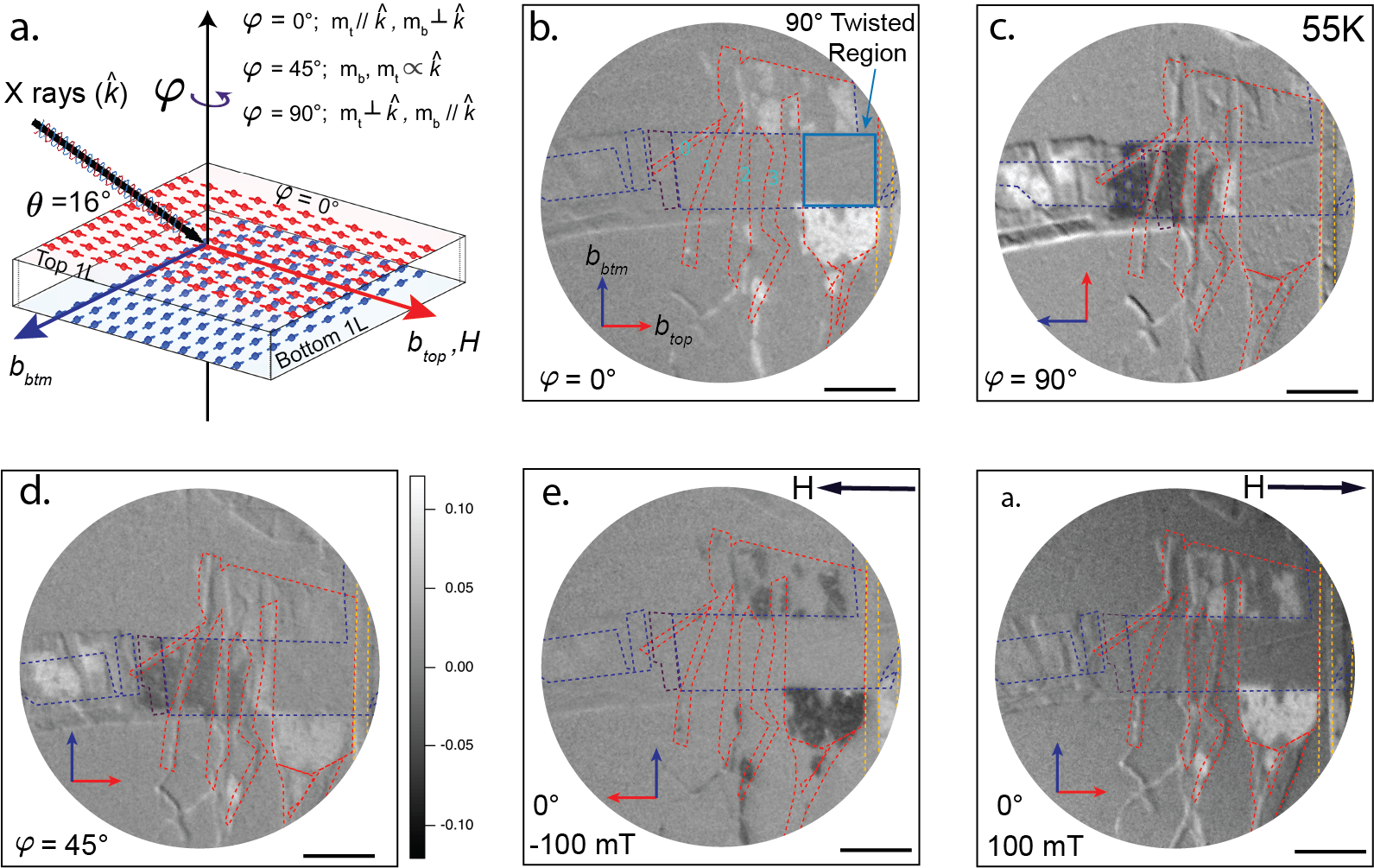}
  \caption{\label{Figure4}
  \textbf{Visualizing spin structure in twisted bilayer CrSBr.}
  \textbf{a}, Schematic of measurement geometry with X-ray beam at $16{^\circ}$ grazing incidence. Red arrows represent top-layer magnetization (along $\vec{b}_{\mathrm{top}}$-axis) and applied magnetic field direction; blue arrows represent bottom-layer magnetization (along $\vec{b}_{\mathrm{bottom}}$-axis, perpendicular to $\vec{b}_{\mathrm{top}}$). The XMCD signal is proportional to $m \cdot \hat{k}$, the projection of magnetization onto the X-ray propagation direction ($\hat{k}$). Colored dashed outlines indicate layer boundaries (red: top layer; blue: bottom layer). The cyan box indicates the $90{^\circ}$ twisted region.
  \textbf{b--d}, XMCD-PEEM maps acquired at azimuthal rotation angles of $\varphi = 0{^\circ}$, $90{^\circ}$, and $45{^\circ}$ (sample rotated counterclockwise about surface normal) under zero applied field at $T = 50$~K.
  \textbf{b}, At $\varphi = 0{^\circ}$, the top layer exhibits strong XMCD contrast with well-defined FM domains. The bottom layer and twisted region display minimal contrast.
  \textbf{c}, At $\varphi = 90{^\circ}$, the bottom layer shows strong XMCD contrast with domains aligned along $\vec{b}_{\mathrm{b}}$, while the top layer ($m_t \perp \hat{k}$) becomes nearly invisible.
  \textbf{d}, At $\varphi = 45{^\circ}$, both top and bottom layers display comparable XMCD signal with visible domain structures, while the twisted region exhibits no significant XMCD contrast. The persistent absence of XMCD signal in twisted regions across all azimuthal angles indicates suppression of in-plane FM order or non-collinear arrangement of spins.
  \textbf{e}, Under applied field $\vec{H} = -$100~mT along $\vec{b}_{\mathrm{top}}$, top layer domains align uniformly with the field direction.
  \textbf{f}, The XMCD contrast in top layer further reverses with 100~mT field reversal field applied, whereas the bottom layer and twisted region remains unaffected by the applied field. Scale bars: 2~$\mu$m. Grayscale colorbars indicate XMCD contrast intensity.}
\end{figure*}

The monolayer (1L) exhibits FM order with all Cr spins aligned identically, yielding net magnetization parallel to the easy axis. In this configuration, XMLD is suppressed because uniform spin alignment produces negligible linear magnetic dichroism, and XLD is primarily dominated by the XNLD contribution from crystal field effects. In multilayers (2L-4L), the weak AFM interlayer coupling stacks successive FM layers antiparallel, producing A-type AFM order. The even-layer (2L, 4L) regions are therefore nearly compensated while, the trilayer (3L)  displays two-fold degeneracy, with net magnetic moment oriented either parallel or antiparallel to the $b$-axis.

Our density functional theory (DFT) calculations  correctly capture the observed layer parity effect, giving positive (FM) intralayer and negative interlayer coupling strengths with magnetic easy axis pointing along crystallographic $b$-axis. The supplemental information outlines the density functional results, Fig.~S4, ~S7 and Table S2. Combined with the thickness dependent XNLD contribution from crystal anisotropy and XMCD arising from Néel axis orientation, XLD intensity across even layers and 3L region significantly exceeds that of a FM monolayer.

To map the response to external in-plane magnetic fields, we apply multiple in-plane magnetic field pulses in the range of $0 -100$ mT on the order few tens of seconds along the easy axis and image the remanent magnetization state after field removal. The observed net magnetization is strikingly layer dependent, odd numbered layers, which possess nonzero net magnetization, reorient in field direction, while even-layer regions, characterized by compensated AFM order, show negligible response (Fig.~2e). At low temperature (45~K), reversing the field direction completely reverses the 1L magnetization (Fig.~2f), while 2L, 3L, and 4L configurations remain largely unchanged, demonstrating their relative field stability.

Furthermore, the response to magnetic field is characterized near the magnetic transition temperature (Fig.~3). We perform field-cooled experiments from room temperature to 110~K in the presence of a static field of $-$25~mT along the easy axis. The XLD-PEEM map reveals finite and uniform linear dichroic contrast across all layers (Fig.~3a). The odd-numbered layers align preferentially along the field-cooling direction; however, a small region in the 3L exhibits magnetization oppositely oriented to the applied field (Fig.~3b), attributed to domain pinning due to impurities. Even layers display negligible XMCD contrast, confirming their compensated AFM state. Application of $-$100~mT saturates the net magnetization in odd layers fully along the field direction (Fig.~3c). The domain structure after subsequent reversal field pulses of +5, +25, and +100~mT are shown in (Fig.~3d–f). At +5~mT, the 1L magnetization nearly reverses, while 3L and 4L remain largely unaffected. Increasing the field to +25~mT completely saturates the 1L along the field direction and partially reverses the 3L, producing two coexisting magnetization orientations within the 3L region. At +100~mT, the 3L is fully saturated along the applied field direction. Across all applied fields, even-layer regions exhibit no detectable magnetization change, consistent with their fully compensated AFM ground state.

The coercive field (field required for magnetization reversal) is significantly reduced for the monolayer at elevated temperatures and is substantially larger for 3L compared to 1L at both temperatures. For even-layer regions, no complete AFM to FM switching transitions are observed across both temperature regimes, corroborating the intrinsic stability of compensated antiferromagnetism in these layers up to 100~mT . Although we note that the small residual magnetization signals in even layer regions at low temperatures could originate from layer by layer switching of spins or structural defects and asymmetric interfaces, but these do not substantially alter compensated AFM behavior \cite{ye2022layer,liu2025spin,krelle2025magnetic,sun2025resolving}. 

\subsection*{Magnetic Ground State in 90$^{\circ}$ Twisted Bilayer CrSBr}

To probe the impact of crystallographic rotation on interlayer magnetic coupling, we investigate a bilayer CrSBr heterostructure (H03) primarily assembled at a $90^{\circ}$ twist angles, where the magnetic easy axis of the top monolayer is orthogonal to the magnetic easy axis of the bottom layer (Fig.~4a). The sample also possesses regions (0,1,2,3) with smaller twist angles of $40^{\circ}, 72 ^{\circ}, 85^{\circ}$ and $81^{\circ} $ respectively. In contrast to the untwisted layers, where every layer retains the crystallographic $b$-axis as its magnetic easy axis, orthogonally twisting two monolayers (red: top; blue: bottom monolayer) produces qualitatively different magnetic ground state in the twisted region (Fig.~4b-d) . 

Since the local XMCD and XLD contrast reflects the projection of the magnetic and orbital moments onto the incident X-ray direction, the observed signal depends sensitively on the sample azimuthal orientation ($\varphi$) with respect to the X-ray direction. Because the in-plane magnetic anisotropy of monolayer CrSBr confines spins predominantly along the crystallographic $b-$axis, the dichroic response varies sharply with azimuthal orientation, providing a means to distinguish the individual monolayer regions from the twisted region. 

The magnetic domain structure of individual layers and twisted regions are mapped by azimuthal angle dependent measurements by rotating the sample plane about its surface normal (Fig.~4b--d). At $\varphi = 0^{\circ}$, i.e., when X-ray propagation is parallel to the top monolayer easy axis, the top layer exhibits a strong XMCD signal, while the bottom layer (with $90^{\circ}$ rotation with respect to the top layer) shows negligible magnetic contrast, since its spins lie along an easy axis nearly perpendicular to the X-ray sensitivity direction and therefore projected weakly onto it. On the other hand, at $\varphi = 90^{\circ}$, the bottom layer exhibits strong XMCD signal while the top layer shows negligible contrast (Fig.~4c). At $\varphi = 45^{\circ}$, top and bottom layers show reduced but comparable magnetic contrast due to their similar projection angles (Fig.~4d). Crucially, the XMCD intensity in the twisted regions remains suppressed at all the three azimuthal angles.

Currently, for the case of orthogonally twisted CrSBr structures, it is understood that either the two layers retain spin orientation  along their respective easy axes or the net magnetization of the twisted region is a superposition of two orthogonal in-plane axes~\cite{boix2024multistep,li2025twist}. In either scenario, the twisted region should exhibit a finite dichroic signal at some azimuthal angle. For instance, if the top (bottom) layer's easy axis were preserved in the twisted region, the twisted region would mirror top (bottom) layer contrast at $\varphi = 0 (90)^{\circ}$ respectively. Instead, we observe negligible dichroic contrast in orthogonally twisted region at $\varphi = 0,90^{\circ}$ (Fig.~4b, c). Furthermore, the absence of XMCD contrast in orthogonally twisted region for $\varphi = 45^{\circ}$ corroborates that neither the net magnetization of individual layers are  preserved nor they exist as superposition of each layer in the twisted region (Fig.~4d).

Additionally, modified interlayer coupling in orthogonally twisted regions can lead to two scenarios. First, a strong interlayer antiferromagnetic coupling can force the two layers to align antiparallel, so that their moments cancel and the N\'eel vector orients along an in-plane direction (but not collinear with individual layers' easy axis), and yielding near-zero net XMCD signal. Secondly, it has been theoretically predicted that 90$^{\circ}$ twist angle can modify (reduce) the interlayer coupling and stabilize a weak ferromagnetic phase with perpendicular magnetic anisotropy in the twisted region, and in our experimental setup this will lead to moments being projected weakly onto the in-plane X-ray direction giving rise to near zero XMCD signal. We rule out the first scenario, i.e., in-plane AFM order with rotated  N\'eel vector, by measuring the XLD-PEEM at two sample azimuthal angles ($\varphi = 45^{\circ}$ and $90^{\circ}$) across regions with varying twist angle between the top and bottom CrSBr layers (Fig.~S10 in the Supplemental Information). The XLD maps and the line profile shows finite XLD in all regions, with spatial variation in XLD intensity between regions of different twist angle. If the N\'eel vector were in-plane, it's projection on the X-ray should change as we vary the sample azimuthal, and by an amount that depends on the twist angle between the top and bottom layer. Instead, as shown in Fig.~S10 (Supplemental Information), the relative ordering of XLD peak intensity across regions of different twist angles remains unchanged under the azimuthal rotation. Since, XLD tracks both local orbital anisotropy and N\'eel axis orientation, the invariant peak structure under rotation points to a structural reconstruction in the twisted region leading to a modified local crystalline anisotropy rather than simple in-plane reorientation of N\'eel vector.

Furthermore, we have theoretically explored the existence of a emergent ground state in the orthogonally twisted interface, by constructing a $4 \times 3 \times 1$ super cell from two orthogonally twisted CrSBr monolayers. The initially orthorhombic cell relaxes into a nearly square lattice under minimal strain to preserve periodicity. Projected density of states (PDOS) analysis further reveals that Br-$p$ and S-$p$ are dominating at valence band maxima (VBM) while Cr-$d$ atoms contributing mostly at conduction band minima (CBM). The magnetic anisotropy energy calculations (MAE) and exchange coupling constants calculations reveal a drastic reduction in the  magnetic anisotropy energy at the twisted region, where competing local contributions from Cr, S, and Br atoms lead to cancellation effects. In agreement with a previous study~\cite{zhang2025twist}, our calculations predict that the magnetic easy axis in twisted region is aligned along the crystallographic $c-$axis. This indicates that the interplay of orthogonal rotational misalignment which reshapes the orbital overlap, lattice geometry, orbital hybridization, and spin orbit driven anisotropy pathways governs the emergence of a new magnetic ground state in the orthogonally twisted region. The magnetic properties and the spin resolved band structure are summarized in Supplemental Material (Table~S2 and Fig.~S8). 

The observed magnetic phase in the twisted region is robust against application of magnetic field (up to $\pm 100$~mT) along the easy axis of the top monolayer (intermediate axis of bottom layer). The $\pm 100$~mT is enough to drive magnetization switching in the top monolayer but leaves the bottom layer and twisted region unchanged (Fig.~4e,f). We observe no new field driven domain wall at the twisted interface or any alignment with the field direction as postulated using atomistic spin dynamic simulations \cite{boix2024multistep}. This observation is consistent with a magnetic ground state that no longer lies along the in-plane easy axis, though we cannot exclude a non-collinear canted configuration at higher fields.

\section*{Conclusion\label{Sonclusion}}

Using element specific XMCD/XMLD-PEEM, we have resolved the spin structure of CrSBr continuously from the multilayer regime to the monolayer limit and into the orthogonally twisted bilayer. In the untwisted films we directly visualize layer-parity-dependent magnetization set by surface termination, consistent with A-type AFM order, and follow its evolution under temperature and in-plane magnetic field. The central result concerns the $90^{\circ}$ twisted interface which deviates from the prevailing picture that individual monolayers retains its in-plane easy axis. Our azimuthal dependent measurements reveal suppressed XMCD contrast at the twisted interface and the rotation invariant XLD rules out an in-plane N\'eel vector. Rather than preserving the monolayer easy axes, the twisted interface undergoes a structural reconstruction that reshapes the local orbital overlap and spin orbit driven anisotropy pathways. Our first-principles calculations corroborate this, showing a collapse of the in-plane magnetic anisotropy at the $90^{\circ}$ twisted region, with the suppressed residual anisotropy energy and the easy axis aligning toward the out-of-plane $c$ direction.
Broadly, our results suggest that large crystallographic twist in CrSBr can qualitatively restructure the magnetic ground state rather than simply superposing the constituent orders. Combined with the layer parity and twist control demonstrated, this positions CrSBr as a versatile platform for designing antiferromagnetic spintronics.

\section*{Methods}\label{Methods}

\subparagraph{Bulk CrSBr Synthesis.}
CrSBr single crystals were grown using the direct solid-vapor method as described 
in Ref.~\cite{liu2022three}.

\subparagraph{Device Fabrication.}
Monolayer graphene was exfoliated onto Si/SiO$_{2}$ substrates (285~nm oxide) using a hot-tape method. CrSBr and few-layer graphite flakes were exfoliated onto Si/SiO$_{2}$ substrates using polydimethylsiloxane (PDMS) stamps in an Ar filled glove box ($<$0.01~ppm O$_{2}$, $<$0.01~ppm H$_{2}$O). All heterostructures (samples H01, H02, and H03) were assembled via sequential pickup using a polycarbonate (PC) film on PDMS: graphene was picked up first, then CrSBr, then graphite, creating a graphene/CrSBr/graphite stack that was transferred onto Pt/Ti/SiO$_{2}$/Si substrates using a custom-built micromanipulator with an optical microscope. The 20~nm Pt layer (with 5~nm Ti adhesion layer) was deposited by e-beam evaporation to provide electrical conductivity for PEEM.

For twisted bilayer fabrication (sample H03), two CrSBr monolayers were sequentially stacked with controlled 90$^{\circ}$ rotation during the pickup process. Twist angle was determined by aligning crystal edges visible in optical microscopy, with estimated accuracy $\pm 1^{\circ}$. All exfoliation and stacking were performed in the Ar-filled glove box. After transfer, residual PC was dissolved in chloroform (15~min), followed by isopropyl alcohol rinse and Ar blow dry inside glove box.

\subparagraph{Atomic Force Microscopy.}
AFM measurements were performed in tapping and contact mode using a Bruker Dimension Icon system with antimony-doped silicon probes ($\mu$masch HQ:NSC15/Al BS, resonant frequency $\sim$325~kHz, spring constant $\sim$40~N/m). Topography images were acquired with $1024 \times 1024$ pixel resolution and analyzed using Gwyddion software  \cite{nevcas2012gwyddion}. Layer thicknesses were determined from height profiles (Supplemental Material, Fig.~S1 and Table~S1).

\subparagraph{PEEM Measurements.}
Synchrotron based XAS and XMCD/XMLD PEEM measurements were performed at the SPEEM end station of the UE49-PGMa beamline at BESSY II synchrotron-radiation facility operated by Helmholtz-Zentrum Berlin (HZB)  \cite{kronast2016speem}. The beamline delivers circularly and linearly polarized soft X-rays with photon energy range 100--2000~eV and resolving power $E/\Delta E > 10^{4}$. For the imaging, photoelectrons were collected with $-10$~kV sample bias, start voltage close to 0 to capture the secondary electrons cascade, and focused by magnetostatic lenses onto a direct, event-counting TimePix3 electron detector. Typical acquisition time was 30--60~s per image with $512 \times 512$ pixel resolution.

Samples were mounted on a six-axis manipulator with temperature control (liquid He cryostat, base temperature 45~K, stability $\pm 0.5$~K) and magnetic field capability (up to $\pm 100$~mT in-plane). X-rays impinged at grazing incidence angle $\theta = 16^{\circ}$ relative to the sample surface. Sample azimuthal rotation ($\varphi$) was controlled via the manipulator rotation stage.

For XAS-PEEM measurements, the spectra was measured by monitoring the intensity of the photoemitted electrons as a function of the photon energy. The presented spectra are divided by the reference current $I_o$ for each photon energy from the incident photon flux (mirror current) to normalize the spectra and the intensity axis is rescaled. Drift correction was performed before selecting the desired region of interest (ROI) for further analysis. XMCD asymmetry was used for spectra and XMCD-PEEM visualization and was calculated as $A_{\mathrm{XMCD}} = (\sigma^{+} - \sigma^{-})/(\sigma^{+} + \sigma^{-}) \times 100\%$, where $\sigma^{\pm}$ denotes intensity measured with right/left circular polarization. XMLD asymmetry was calculated similarly using linear horizontal/vertical polarization.

For XMCD measurements on untwisted samples, the incident X-rays were aligned parallel to the magnetic easy axis of CrSBr ($\varphi = 0^{\circ}$), maximizing sensitivity to the magnetization component along the easy axis. For XMLD measurements on untwisted samples, the incident X-rays were aligned perpendicular to the magnetic easy axis ($\varphi = 90^{\circ}$), enabling detection of magnetic anisotropy and crystal field effects. For twisted samples, azimuthal-angle-dependent XMCD and XMLD measurements were performed by rotating the sample about its surface normal through various angles, allowing direct observation of how magnetic contrast varies with sample orientation.

For XMCD-PEEM imaging, images were acquired at fixed photon energies corresponding to maximum dichroic contrast (Cr $L_{3}$ peak $\sim$576~eV). XMCD-PEEM images were generated by following the XMCD asymmetry as defined above.

As PEEM is intrinsically a surface sensitive technique, with typical inelastic mean free path of 2--3 nm, the measured intensity therefore represents a depth-dependent, exponentially attenuated sum of contributions from all layers. The contributions from depth $z$ can be expressed approximately as $\exp(-z/\lambda)$, where $\lambda$ is the effective escape depth. The dichroic signal can thus be expressed as:
\begin{equation*}
   S \propto \sum_i M_i \exp(-z_i/\lambda),
\end{equation*}
with $M_{i}$ the magnetization of layer $i$ at depth $z_{i}$. In antiferromagnetically coupled multilayers with alternating magnetization, this attenuation prevents complete cancellation of oppositely aligned layers, yielding a finite net XMCD contrast for an ideally compensated AFM stack. Successive layers contribute with progressively reduced weight, such that the first layer contributes more strongly than the second, the third more than the fourth, and so forth. This unequal depth weighting yields a finite net XMCD contrast even for an ideally compensated AFM stack, such that XMCD--PEEM probes a depth-weighted magnetic imbalance rather than a purely surface magnetization.

\subparagraph{Computational Methods.}
Geometry optimization and electronic structure (spin-polarized) ab-initio calculations were performed by using DFT~\cite{hohenberg1964inhomogeneous,kohn1965self}, implemented in the Vienna ab initio simulation package (VASP) \cite{kresse1996efficient}. Exchange correlation energies are treated according to generalized gradient approximation (GGA), employing the Perdew-Burke-Ernzerhof (PBE) parameterization \cite{perdew1996generalized,perdew1998perdew,hammer1999improved}. The long range vdW interactions were incorporated using Grimme's DFT-D3 dispersion correction with Becke-Johnson (BJ) damping (IVDW = 12) throughout all calculations \cite{grimme2011effect,grimme2010consistent}. The electron-ion interactions are approximated by projector augmented wave (PAW) method \cite{pe1994projector} with  ($s^{1}$, $p^6$ and $d^{5}$) states for Cr,  ($s^{2}$ and $p^{4}$) for  S atoms, and ($s^{2}$ and $p^{5}$) states for Br being treated as valence states, using plane wave energy cutoff of 450 eV.
All atomic structures were fully relaxed until Hellmann-Feynman force on each atom is reached at the threshold of  -0.02 eV {\AA}$^{-1}$ and total energies are optimized until the energy difference between successive iteration becomes $10^{-7}$ eV.
Brillouin zone (BZ) integration were carried out by sampling the Monkhorst-Pack mesh with $\Gamma$ -centered k-point grids of $12 \times  12 \times  1$ for structural relaxation and $19 \times  19\times  1$ for electronic properties of monolayer and $8 \times  8 \times  1$ for structural relaxation and $11 \times  11\times  1$ for electronic properties of pristine multilayers using $2 \times  2 \times  1$ supercell. A thick layer of 25 {\AA} in the out-of-plane direction was utilized to prevent the self interactions due to Born-von Karman boundary conditions. To describe the on-site Coulomb repulsion of Cr-3d electrons, the LDA + U method has been considered with with U = 4 eV and J = 1 eV, corresponding to an effective Hubbard parameter $U_{eff}=3$ eV \cite{anisimov1991band}. 
For the MAE calculations, the spin-orbit coupling (SOC) was taken into account. Heisenberg exchange interaction (HSI), and
 MAE were calculated via the QuantumATK-Synopsys package, employing an LCAO basis set, the "PseudoDojo" pseudopotential \cite{smidstrup2020quantumatk} , a density mesh cutoff of 140 Hartree, and a $17 \times  17 \times  1$ k-point grid.

\paragraph*{Acknowledgments.}
J.K. acknowledges the financial support from U.S. Department of Energy, Office of Science, Office of Basic Sciences, through award No. DE-SC002549 (for measurements, device fabrication and postdoctoral support); from U.S. Office of Naval Research (ONR) under Award No. N00014-23-1-2751; the Center for Emergent Materials at The Ohio State University, an National Science Foundation (NSF) MRSEC, through Award No. DMR-2011876; and from NSF-CAREER Award under Grant No. DMR-2339309.
S.S. acknowledges the financial support from National Science Foundation (NSF) through Grants No. DMR-2210510, ECCS-2531211, and from the Center for Emergent Materials at The Ohio State University, an NSF MRSEC, through Award No. DMR-2011876; ONR under Award No. N00014-23-1-2751. S.S. also acknowledges financial support from NSF-CAREER Award through Grant No. ECCS-2339723. The work at UT Dallas is supported by NSF 2516364, ONR N00014-23-2020 and AFOSR FA9550-19-1-0037. A.X.G. acknowledges support from the U.S. Department of Energy, Office of Science, Office of Basic Energy Sciences, Materials Sciences and Engineering Division under Award No. DE-SC0024132, as well as support from the Alexander von Humboldt Foundation. The authors thank the Helmholtz--Zentrum Berlin for the provision of access to synchrotron radiation facility and allocation of synchrotron radiation at the SPEEM end-station of UE49-PGMa beamline of BESSY II at HZB. The authors also acknowledge the use of the Claire and John Bertucci Nanotechnology Laboratory at Carnegie Mellon University for the deposition and patterning of Pt/Ti on SiO$_2$/Si substrates. 

\paragraph*{Author Contributions.}
J.K.\ and S.S.\ conceived the project, designed the experiments, and supervised the project. R.K.B., Z.C., R.P. and A.T.\ fabricated the heterostructures of CrSBr. I.H.K., A.T., and S.P.\ performed AFM measurements. A.T., S.P., A.S.,  J.K., F.K., A.X.G.,and S.S.\ performed XMCD/XMLD-PEEM measurements. A.T. analyzed the data, with assistance from S.P. and S.A.E.. W.L., A.L.N.K\ and B.L.\ grew the CrSBr bulk crystals. A.K. performed density functional calculations under the supervision of V.S.  B.T.\ helped to prepare the illustration shown in Fig.~1. All authors contributed to the scientific discussion of the results. A.T. wrote the manuscript with input from all the authors.

\paragraph*{Competing Interests.}
The authors declare no competing interests.

\bibliographystyle{apsrev4-2}
\bibliography{Manuscript/references}

\clearpage
\newpage

\setcounter{figure}{0}
\setcounter{table}{0}
\setcounter{equation}{0}

\renewcommand{\thefigure}{S\arabic{figure}}
\renewcommand{\thetable}{S\arabic{table}}
\renewcommand{\theequation}{S\arabic{equation}}
\renewcommand{\thepage}{S\arabic{page}}

\setcounter{page}{1}
\onecolumngrid

\section*{Supplemental Material for: Element-Specific Visualization of Layer-Parity and Twist-Dependent Magnetism in CrSBr}

\section*{Optical and atomic force microscopy of heterostructures}
\begin{figure*}[ht]
\centering
\includegraphics[width=0.75\textwidth]{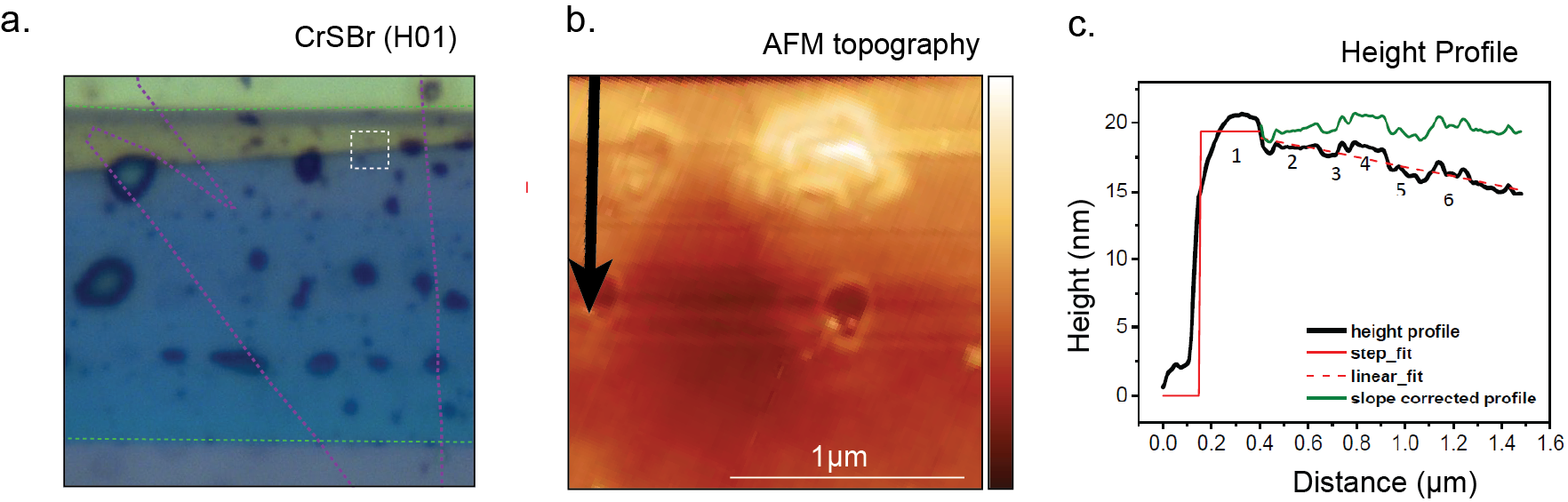}
\caption{\label{multilayerAFM}
\textbf{Optical and atomic force microscopy on multilayer CrSBr heterostructure (H01).}
\textbf{a}, Optical micrograph of the CrSBr heterostructure (sample H01), showing both graphene-encapsulated and bare regions.
\textbf{b}, AFM topography map of the multilayer CrSBr terrace structure. Red arrow indicates the line profile direction.
\textbf{c}, AFM height profile across the terrace steps shown in \textbf{b}, including step-fit, linear-fit, and slope-corrected profiles. Terrace numbers correspond to approximate layer counts from  Supplemental Material Table~S1.}
\end{figure*}

\begin{table*}[hbt!]
\centering
\caption{Layer thickness variation in multilayer CrSBr heterostructure (sample H01). 
Thickness values determined from AFM height profile shown in Supplemental Material
Fig.~S1c. Layer parity (odd/even) determines magnetic behavior as discussed in 
the main text.}
\begin{tabular}{ccc}
  \hline
  \hline
  Terrace& Number of layers & Layer parity \\
  \hline
  1 & 31 & Odd \\
  2 & 28 & Even \\
  3 & 27 & Odd \\
  4 & 28 & Even \\
  5 & 25 & Odd \\
  6 & 26 & Even \\
  \hline
  \hline
\end{tabular}
\label{SI Table:layerinfo}
\end{table*}

\begin{figure*}[ht]
\centering
\includegraphics[width=0.85\textwidth]{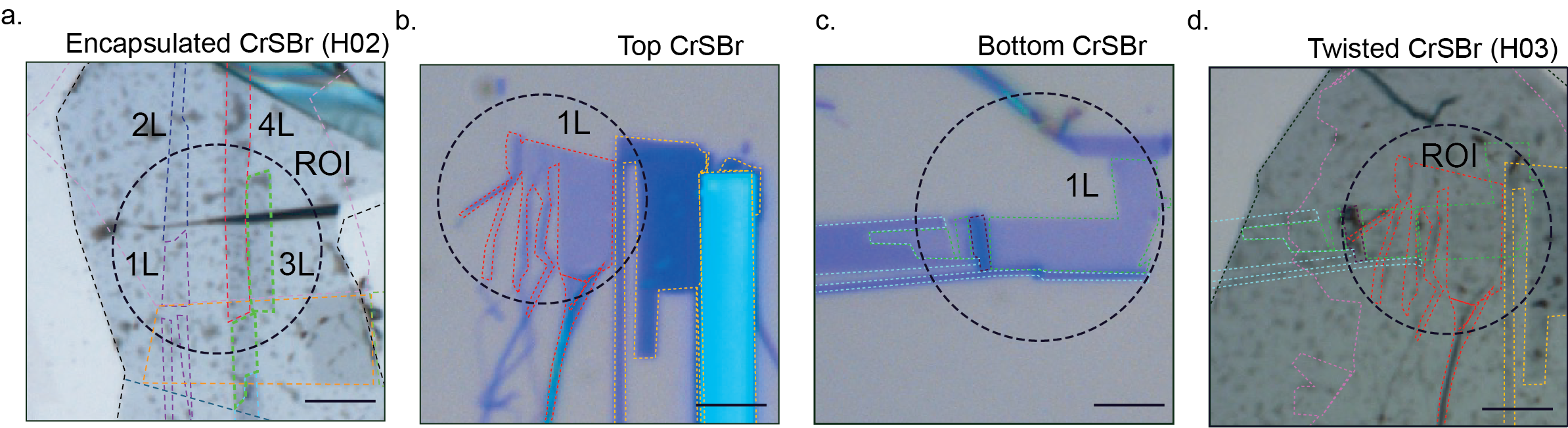}
\caption{
\textbf{Optical micrographs of atomically thin CrSBr heterostructures (H02 and H03).}
\textbf{a}, Optical micrograph of graphene-encapsulated CrSBr heterostructure (sample H02), highlighting the 1L--4L atomically thin flakes. The dashed circle indicates the XMCD-PEEM measurement region shown in main text Fig.~2.
\textbf{b}, Optical image of twisted bilayer CrSBr heterostructure (sample H03). The PEEM measurement region is indicated by the dashed circle.
\textbf{c,d}, Optical images showing the top and bottom monolayer CrSBr used for the twisted heterostructure. Layer thicknesses were confirmed using optical contrast combined with AFM topography measurements. Scale bars: 5~$\mu$m.}
\label{S2}
\end{figure*}

\newpage
\section*{Density functional calculations}
The magnetic exchange interactions are modeled  using a real space Green's function implementation of the original Liechtenstein-Katsnelson-Antropov-Gubanov (LKAG) formula \cite{liechtenstein1987local}, as implemented in QuantumATK \cite{smidstrup2020quantumatk}. The type and the strength of magnetic exchange coupling are employed using Heisenberg model:
 \begin{equation}
H = - \sum_{i \neq j} J_{ij}\,\hat{e}_i \cdot \hat{e}_j,
\label{eq:heisenberg}
\end{equation}
where $\hat{e}_i$ is the normalized local spin vector on atom $i$, $J_{ij}$ is the distance dependent Heisenberg exchange coupling constant. This was achieved by a real-space Green's function implementation of the original LKAG formula:
\begin{equation}
J_{0i,\mathbf{R}j}
=
-\frac{1}{4\pi}
\int_{-\infty}^{E_F}
d\epsilon\,
\mathrm{Im}\,
\mathrm{Tr}
\left[
\Delta_i\,
G_{ij}^{\uparrow}(\epsilon,\mathbf{R})\,
\Delta_j\,
G_{ji}^{\downarrow}(\epsilon,-\mathbf{R})
\right],
\label{eq:lkag}
\end{equation} 
where $i$ and $j$ label atomic indices within a unit cell. Here, $J_{0i,\mathbf{R}j}$ represents the exchange interaction between atom $i$ in the reference unit cell and atom $j$ in the unit cell translated by the lattice vector $\mathbf{R}$. These exchange parameters correspond to the distance-dependent Heisenberg exchange constants $J_{ij}$ used in Eq.~(S1), and \[
\Delta_i = H_i^{\uparrow} - H_i^{\downarrow}\] is the on-site difference between the up and down part of the Hamiltonian matrix. The distance dependent $|J_{ij}|$ diminishes at longer range interactions, and the AFM magnetic exchanges are always predominant.
The magnetic exchange interactions were calculated using the Magnetic Exchange Analysis (MEA) implementation in QuantumATK \cite{smidstrup2020quantumatk}. This method determines the exchange coupling constants within the infinitesimal spin-rotation (magnetic force theorem) formalism \cite{liechtenstein1987local}. The conventional total-energy mapping approach requires multiple magnetic configurations to parameterize a Heisenberg Hamiltonian, whereas the MEA method extracts the exchange interactions directly from the converged electronic structure. The MEA method is computationally efficient for the large twisted supercells considered in the present work, where the number of atoms and possible magnetic configurations increases substantially.
\label{sec:S1}
\begin{figure*}[h]
\centering
\includegraphics[width=0.75\textwidth]{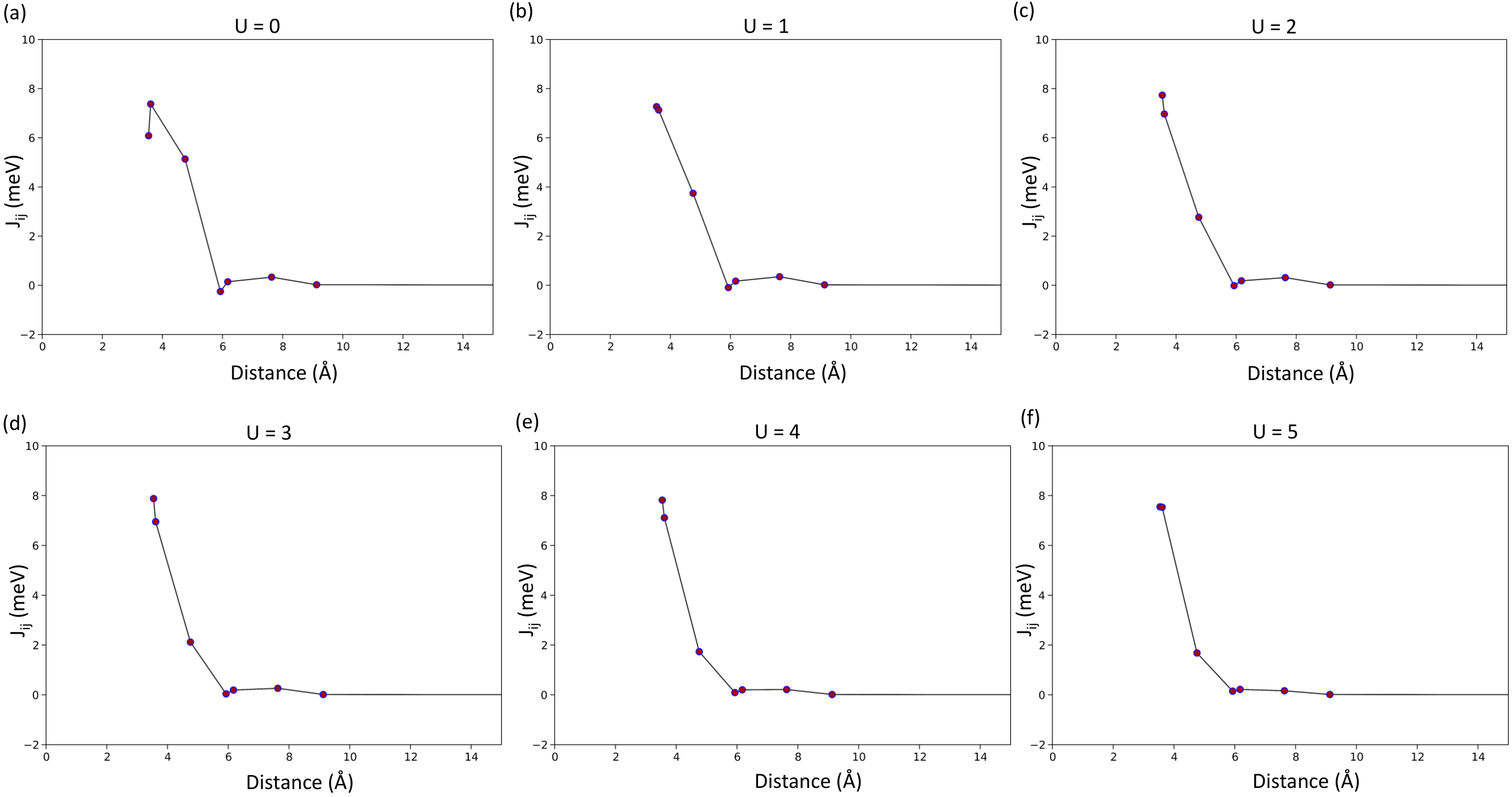}
\caption{Variation of the exchange coupling parameters as a function of interatomic distance for the monolayer CrSBr system at different on-site Coulomb interaction strengths ($U$). Figures (a) to (f) corresponds to $U = 0$, 1, 2, 3, 4, and 5 eV, respectively.}
\label{fig:figuresSVIII}
\end{figure*}
Fig.~\ref{fig:figuresSVIII} summarizes the dependence of the exchange coupling constants on the Hubbard parameter ($U=0$ - 5~eV) calculated within the MEA framework. This reveals that the relative magnitudes of the nearest-neighbour exchange interactions evolve systematically with increasing $U$. At $U=0$, the calculated exchange hierarchy is $J_{2}>J_{1}$, whereas increasing the Hubbard parameter leads to a crossover to $J_{1}>J_{2}$. Independent calculations performed using VASP \cite{kresse1996efficient} retain the hierarchy $J_{2}>J_{1}$ for the reference monolayer and also supported by the literature \cite{guo2018chromium}. Such differences are not unexpected. Although both QuantumATK and VASP employ the rotationally invariant DFT+$U$ formalism, the effective strength of the Hubbard correction depends on the underlying basis set, projector functions, construction of the occupation matrix, and other implementation-specific details. As a result, the same nominal value of the Hubbard parameter can lead to different electronic structures and magnetic exchange interactions in the two codes.
The monolayer calculations therefore serve as a benchmark for establishing an internally consistent protocol within the QuantumATK MEA framework. Since the $U=0$ calculations reproduce the exchange hierarchy obtained from our benchmark VASP calculations for the reference system, this parameterization was used consistently throughout the work for the bilayer and twisted CrSBr bilayer systems. Maintaining the same computational protocol for all structures provides a meaningful comparison of the exchange interactions across the stacked and twisted bilayers while retaining the computational efficiency required for large supercells.
\begin{figure*}
    \centering
    \includegraphics[width=4in]{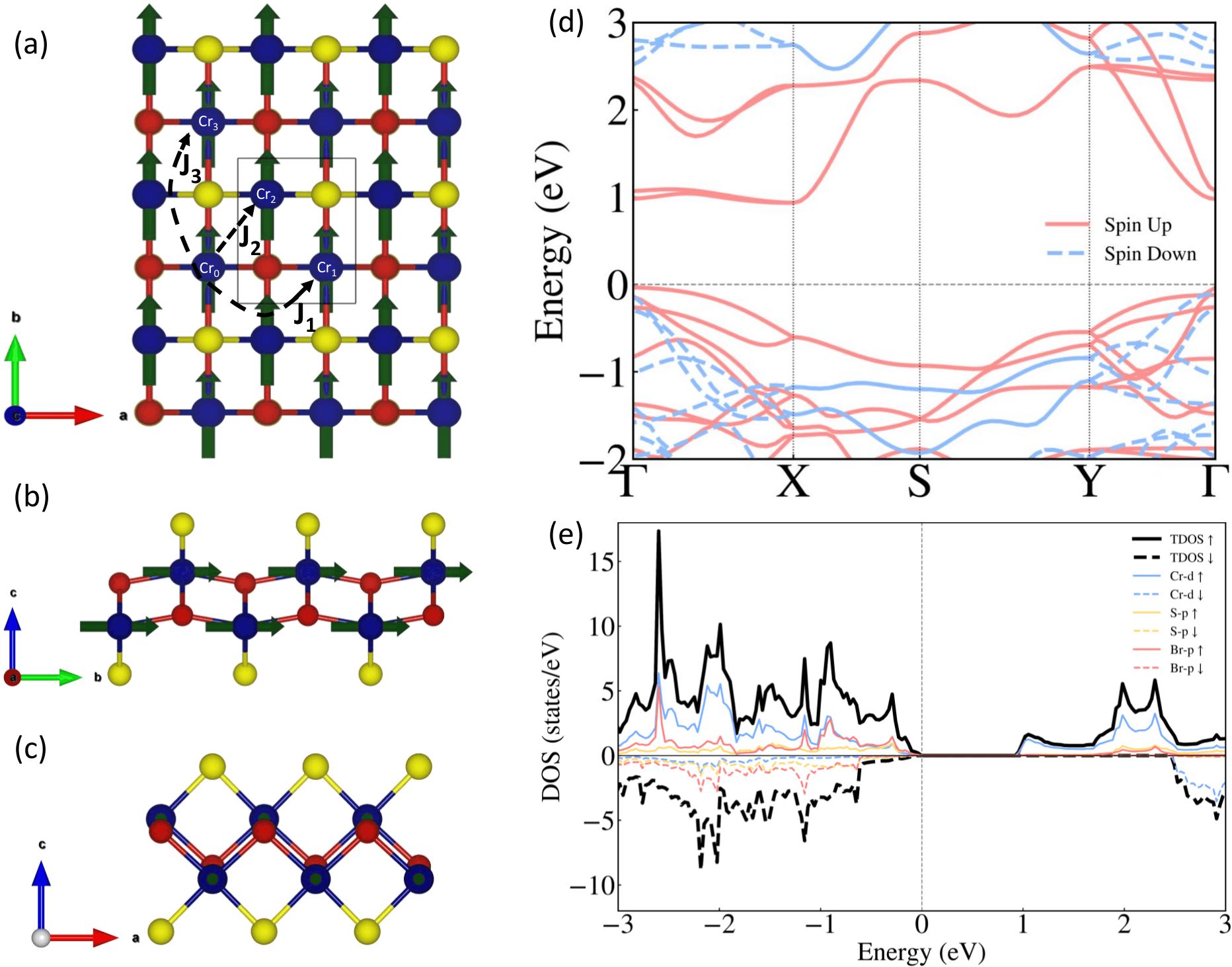}
    \caption{(a) Side and (b, c) top views of the monolayer CrSBr. The grey rectangle in the top figure denote the unit cell. (d) Spin resolved electronic band structure  and (e) projected density of states, the Fermi level is set at zero, and the orbital-resolved contributions are represented by different colors (blue (Cr-d), red (S-p),and yellow (Br-p), whereas the size of the markers denotes the relative orbital weight. }
    \label{DFT_1L} 
\end{figure*}
Since, the dominant mechanism of magnetic exchange coupling in monolayer CrSBr is super-exchange interaction \cite{lee2021magnetic, wang2023magnetic}, the orbital hybridization between Chromium (\textit{Cr}) and anion atoms are crucial to understand the exchange interaction. From the electronic band structure and PDOS analysis, the energy bands below the Fermi level are composed of \textit{Cr-d} orbitals with significant contributions from Sulfur(\textit{S})  and Bromine (\textit{Br}) \textit{p-}  orbitals, while the bands above the Fermi level are primarily composed of \textit{Cr-d} orbitals, exhibiting mixed but significant \textit{Cr}- anion hybridization. The electronic band structure along high symmetry path $\Gamma$ - X - S - Y - $\Gamma$ of the BZ and corresponding projected density of states (PDOS) are calculated and shown in supplementary Fig.~\ref{DFT_1L} for monolayer.  The small \textit{p-d} hybridization induced orbital overlap is nearly isotropic and does not contribute to the magnetic anisotropy. Changing the spin orientation will have an extremely weak influence on the exchange interaction. We find that the exchange constants of monolayer CrSBr only changes slightly compared with those of bulk CrSBr, but following the trend as in the pervious results \cite{scheie2022spin}.
We further calculate the in plane MAE  directly from non collinear self-consistent calculations including spin-orbit coupling which directly related to our imaging results. We used Quantum ATK package to calculate MAE by using the force theorem, which allows the evaluation of the energy difference from non-self-consistent band energies \cite{blonski2009density, blanco2019validity}:
\begin{equation}
\mathrm{MAE}
=
\sum_{i}
f_{i}(\theta_{1},\phi_{1})
\,\varepsilon_{i}(\theta_{1},\phi_{1})
-
\sum_{i}
f_{i}(\theta_{0},\phi_{0})
\,\varepsilon_{i}(\theta_{0},\phi_{0}),
\end{equation}
where the spin orientation is described by the two spherical angles $\theta$ and $\phi$ .The $f_{i}(\theta,\phi)$ is the occupation factor for band $i$ (including both band and $k$-point indices), and $\varepsilon_{i}(\theta,\phi)$ is the corresponding band energy. The details of the calculated exchange spin/orbital moments, coupling strengths, magnetic anisotropy energy, preferred magnetization axis, band gap and interlayer distances  are summarized in Table~\ref{tab:table1} while the variation of exchange coupling constants as a function of layer number and twist angle are shown in Fig. \ref{fig:figuresSI}.  
\begin{figure*}[t]
\centering
\includegraphics[width=0.75\textwidth]{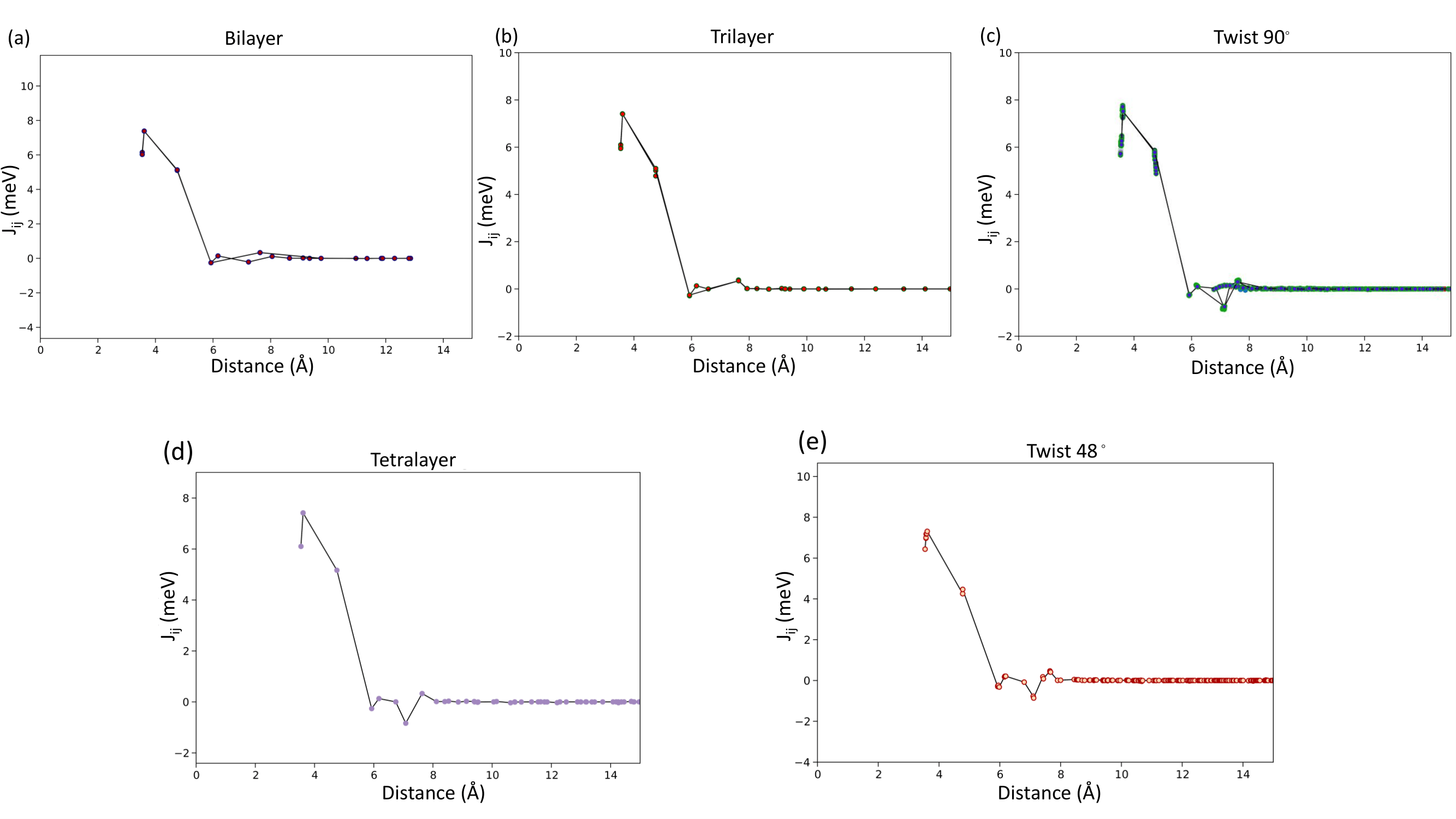}
\caption{Variation of exchange coupling constants for (a) bilayer, (b) trilayer, (c) \ang{90} twist, (d) tertalayer, and (e) \ang{48} degree twisted CrSBr.}
\label{fig:figuresSI}
\end{figure*}
From the atom-resolved MAE analysis shown in Fig.~\ref{fig:figuresSX}(a-f), a clear layer- and structure- dependent redistribution of magnetic anisotropy contributions among Cr, S, and Br atoms is observed in CrSBr. In the pristine monolayer as shown in Fig.~\ref{fig:figuresSX}(a), the MAE is predominantly governed by the Br atoms, while the Cr and S atoms provide comparatively smaller contributions (already explained in the main text). This indicates that the heavy Br atoms, through stronger spin-orbit coupling, play the leading role in stabilizing the in-plane magnetic anisotropy. A similar trend persists in the bilayer system, Fig.~\ref{fig:figuresSX}(b), where Br atoms continue to dominate the local MAE contributions, yielding a total MAE of 2.08 meV for the $2 \times 2 \times 1$ supercell (Cr$_{16}$S$_{16}$Br$_{16}$). This suggests that interlayer coupling does not significantly suppress the Br-driven anisotropy, although a slight redistribution between Cr and S contributions begins to emerge.
\begin{figure*}[h]
\centering
\includegraphics[width=0.75\textwidth]{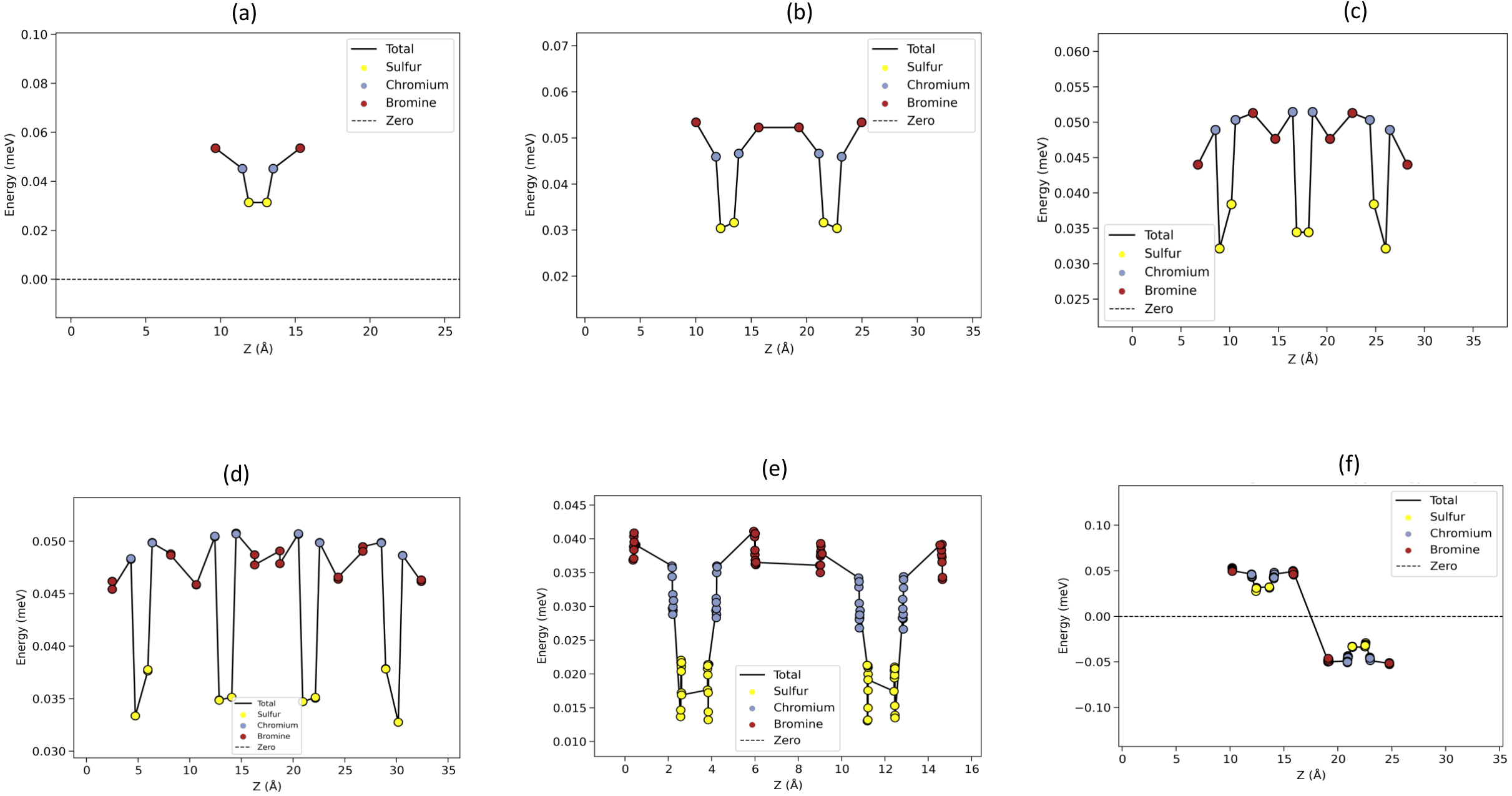}
\caption{Variation of atom resolved MAE  for (a) monolayer, (b) bilayer, (c) trilayer, (d) tetralayer, (e) \ang{48} twisted bilayer, and (f) \ang{90} twisted bilayer CrSBr.}
\label{fig:figuresSX}
\end{figure*}

In the trilayer system as presented in Fig.~\ref{fig:figuresSX}(c), a more cooperative behavior is observed, where both Cr and Br atoms contribute substantially to the MAE. This indicates that increasing layer thickness enhances the participation of Cr-$3d$ states in addition to Br-$p$ states, resulting in an increased total MAE of 3.19 meV. For the tetralayer system, Fig.~\ref{fig:figuresSX}(d), this cooperative enhancement becomes more pronounced, with all three atomic species contributing more uniformly across the layers. The total MAE further increases to 4.24 meV, corresponding to a reduced value of 0.1325 meV/Cr, indicating that the anisotropy energy is distributed over a larger number of magnetic centers.

A significant modification is observed in the twisted bilayer with \ang{48} rotation, Fig.~\ref{fig:figuresSX}(e), where the moir\'e-induced structural reconstruction leads to strong spatial modulation of atomic contributions. Cr atoms exhibit enhanced local fluctuations, S atoms contribute significantly in regions of stronger hybridization, and Br atoms still provide a substantial but spatially modulated contribution. The overall MAE is 3.12 meV, corresponding to 0.086 meV/Cr, indicating partial suppression due to structural frustration and modified interlayer registry.
In contrast, the \ang{90} twisted bilayer as shown in Fig.~\ref{fig:figuresSX}(f) exhibits a drastic change in magnetic anisotropy, where competing local contributions from Cr, S, and Br atoms lead to cancellation effects and a reversal of the MAE (approximately $-0.11$ meV per Cr), despite an overall energy scale of 2.291 $\mu$eV/Cr. This indicates that strong rotational misalignment fundamentally alters orbital overlap and spin-orbit-driven anisotropy pathways.
Overall, the atom-resolved trends demonstrate that Br atoms consistently act as the primary source of spin-orbit-induced anisotropy, while Cr atoms increasingly contribute with higher layer number and structural complexity, and S atoms mainly mediate hybridization effects.

\begin{figure*}[h]
    \centering
    \includegraphics[width=5in]{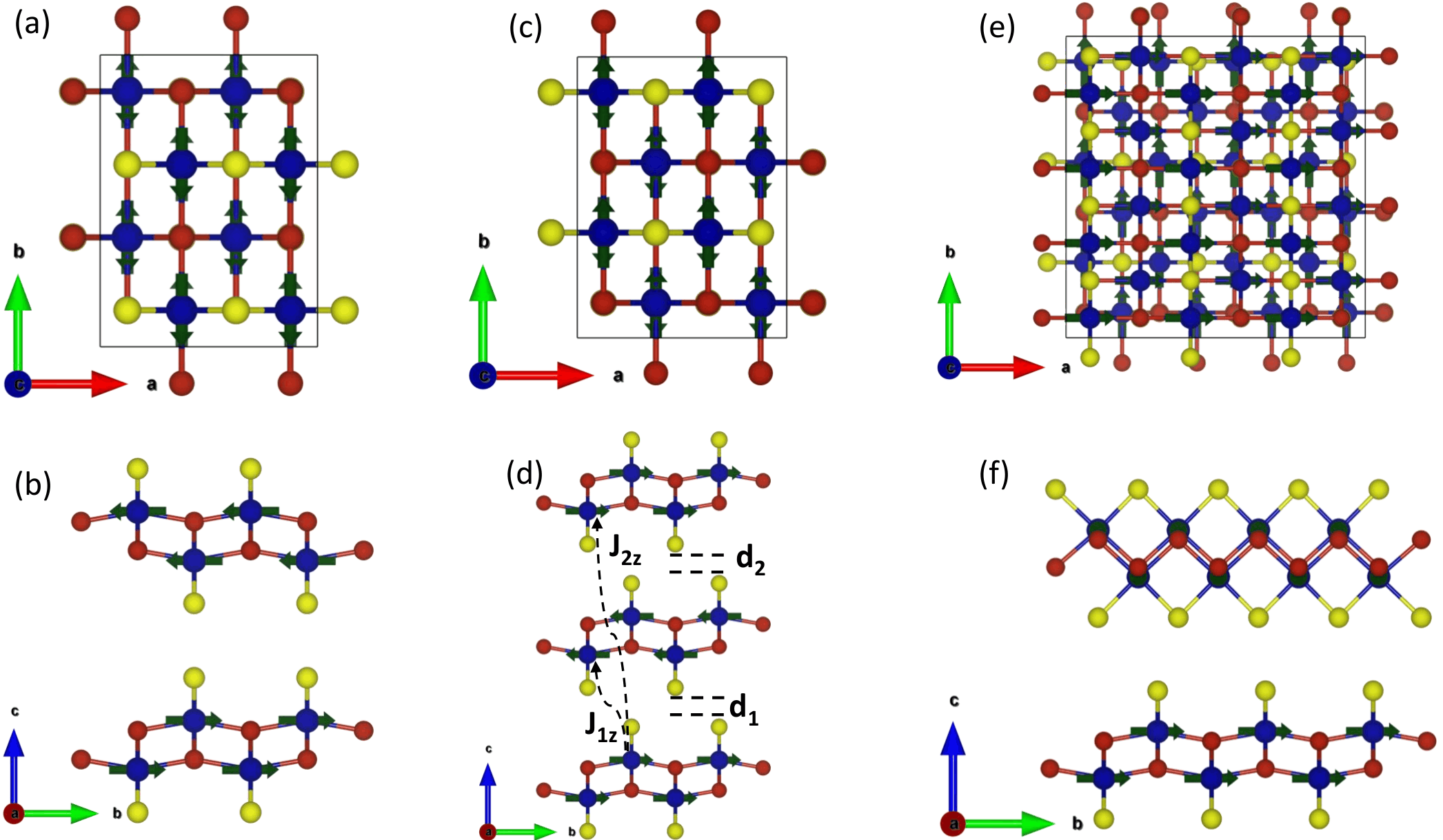}
    \caption{(a, c, and d) top views and (b, d, and f) side views of the bilayer, trilayer and \ang{90} twisted CrSBr respectively. The grey rectangle in the top figure denote the unit cell.}
    \label{fig:Figure 2} 
\end{figure*}

\begin{figure*}[h]
    \centering
    \includegraphics[width=5in]{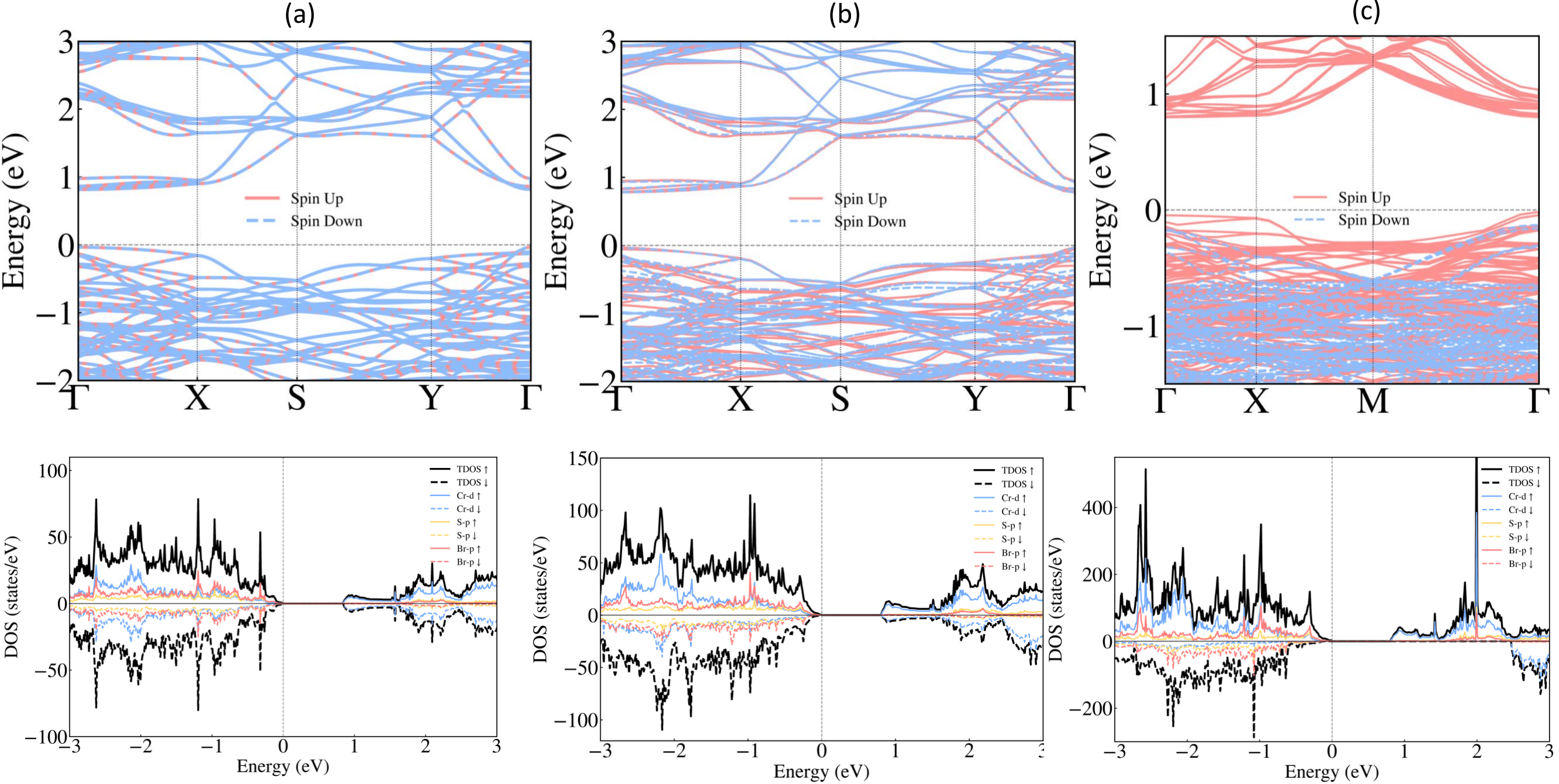}
    \caption{Electronic structures (spin resolved bands, and PDOS) of (a) bilayer, (b) trilayer, and (c) \ang{90} twisted CrSBr. The Fermi level is set at zero, and the orbital-resolved contributions are represented by different colors (blue (Cr-d), yellow (S-p), and red (Br-p), whereas the size of the markers denotes the relative orbital weight.}
    \label{fig:Figure 3} 
\end{figure*}

 The optimized ground state magnetic configuration is shown in Fig.~\ref{fig:Figure 2}(a and b). We take $2 \times  2\times  1$ supercell with the fully relaxed lattice parameters a = 7.08 {\AA} and b = 9.51 {\AA} and interlayer spacing $d_1$ as 3.62 {\AA}. Both ferromagnetic (FM) and various antiferromagnetic (AFM) spin configurations were considered during structural relaxation. The total energy calculations revealed that AFM1 state was the lowest in  energy as shown in the Fig.~\ref{fig:Figure 2}(a and b), also seen in the Table I of  Supplemental Material. It exhibits intralayer magnetic order as FM and interlayer magnetic state is AFM which matches well with previous studies \cite{liu2024intralayer}. The easy axis of magnetization is along b-axis.
Fig.~\ref{fig:Figure 3}(a) shows the projected band structure and PDOS for bilayer CrSBr, revealing it to be an indirect band gap semiconductor with a value of 0.84 eV and very close to the monolayer. This matches with the previous theoretical studies \cite{lee2021magnetic}.  The conduction band minima (CBM) is located at the X point and is mainly contributed by Cr-$d$ orbitals, while the valence band maxims (VBM) is located at the $\Gamma$ point and is mainly contributed by the S-$p$ and Br-$p$ orbitals. Furthermore, it can be seen from the PDOS Fig.~\ref{fig:Figure 3}(a) that the Br-$p$ orbitals contribute most to the lower valence bands. The net spin magnetic moment is $3.205\,\mu_B$/Cr atom as shown in Table~\ref{tab:table1}. It should be noted that the way of stacking  and magnetic type is chosen cautiously because the energy differences among various states are relatively  small, generally in order of a few meV. 

Thus, for each trilayer CrSBr system, the stacking and magnetic configurations dependence becomes obvious. For our convenience, we made different stacking and magnetic orders as shown in Supplemental Material Fig. 5(a - f). The best possible configuration is shown in the Fig.~\ref{fig:Figure 2}(c and d). As inherited from bilayer, the intraplanar magnetic state is FM while interplanar magnetic ground state is AFM to the alternate layers followed by b - axis as the easy axis of magnetization. In Fig.~\ref{fig:Figure 3}(b) we showed the spin resolved electronic band structure for the trilayer CrSBr, where VBM and CBM are mostly contributed from spin-up and spin-down states equally although net magnetization is not zero, followed by semiconducting direct band gap at $\Gamma$ point with a value of 0.78 eV. From PDOS, the VBM is mostly contributed from Cr-$d$, S-$p$, and Br-$p$ orbitals but CBM is dominantly contributed by Cr-$d$ orbitals. The magnetic properties obtained from the calculations are shown in Table~\ref{tab:table1}.

\begin{table*}[h]
\centering
\caption{Summary of magnetic properties for monolayer, bilayer, trilayer, and 90$^\circ$ twisted structures, including spin and orbital magnetic moments per Cr atom ($\mu_{\mathrm{spin}}/\mathrm{Cr}$, $\mu_{\mathrm{orb}}/\mathrm{Cr}$), exchange parameters ($J_1$--$J_{2z}$ in meV), and magnetic anisotropy energy (MAE in meV/Cr).}
\footnotesize
\setlength{\tabcolsep}{6pt}
\begin{tabular}{c cc ccccc c c } 
\hline
\hline
System &
$\mu_{\mathrm{spin}}$ &
$\mu_{\mathrm{orb}}$ &
$J_1$ &
$J_2$ &
$J_3$ &
$J_{1z}$ &
MAE &
Easy Axis &
$\Delta (eV)$ \\
\hline
Monolayer      & 3.115 & 0.018 & 6.08 & 7.37 & 5.13 & --     & 0.130 & b & 0.96 \\
Bilayer        & 3.205 & 0.016  & 6.03 & 7.39 & 5.13 & -0.26 & 0.130 & b & 0.84 \\
Trilayer       & 3.210  & 0.015  & 5.94 & 7.40 & 5.10 & -0.26 & 0.133 & b & 0.78 \\
Tetralayer       & 3.120  & 0.018 & 5.95 & 7.41  & 5.24 & -0.25   & 0.133 & b & 0.78 \\
48$^\circ$ Twist & 3.115 & 0.016 & 6.57 &  7.56   &   7.19   &    0.13   & 0.086 & c & 0.85 \\
90$^\circ$ Twist & 3.209 & 0.015 & 6.17 &  7.48   &   7.33   &  0.12     & 0.002 & c & 0.81 \\
\hline
\hline
\end{tabular}
\label{tab:table1}
\end{table*}

\clearpage
\newpage
\section*{Twist angle dependent XLD-PEEM map at 45~K in H03}

\begin{figure}[h]
  \centering
  \includegraphics[width=0.9\textwidth]{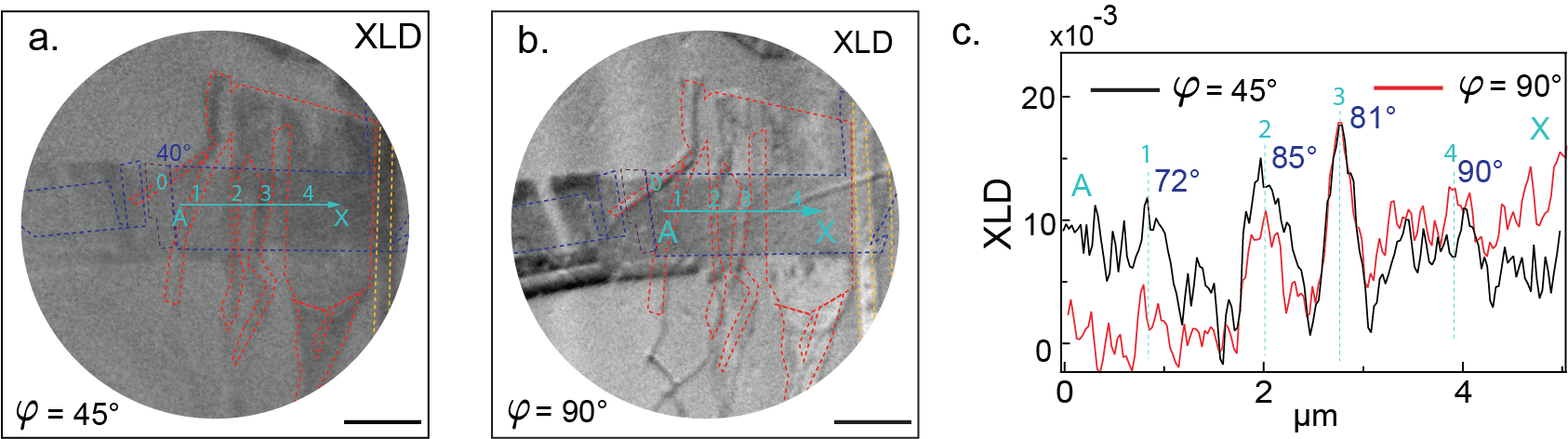}
  \caption{\label{S5}
  \textbf{ XLD-PEEM line profile as a function of twist angle between top and bottom monolayer}
  \textbf{a}, XLD-PEEM map at $\varphi = 45^{\circ}$ azimuthal angle showing finite linear dichroic contrast in both twisted and non-twisted regions.
  \textbf{b}, XLD-PEEM map showing finite linear dichroism across the sample for X-rays aligned orthogonal to the b-axis of the top layer. Note that the top and bottom layers have b-axes separated by $90^{\circ}$. Scale bar: 2~$\mu$m.
  \textbf{c}, Raw XLD line profile extracted along the marked direction showing similar relative intensity when the sample azimuthal angle is $45^{\circ}$ or $90^{\circ}$ relative to the X-ray direction. The systematic variation in XLD with twist angle and independent of sample azimuthal is indicative of a structural reconstruction at the twisted interface potentially driven by superlattice modulation rather than potential AFM order with varying Nèel axis. The data was acquired at 45~K}
\end{figure}

\end{document}